\documentstyle[twocolumn,aps,epsf]{revtex}
\begin{document}
\newcommand{\beq}{\begin{equation}}
\newcommand{\eeq}{\end{equation}}

\title{SECOND HARMONICS AND COMPENSATION EFFECT IN CERAMIC SUPERCONDUCTORS}

\author{Mai Suan Li}

\address{Institute of Physics, Polish Academy of Sciences,
Al. Lotnikow 32/46, 02-668 Warsaw, Poland }

\address{
\centering{
\medskip\em
{}~\\
\begin{minipage}{14cm}
A three-dimensional lattice
of the Josephson junctions with a finite self-conductance is employed
to model the ceramic superconductors.
The nonlinear ac susceptibility and the compensation effect 
are studied by Monte Carlo simulations in this model.
The compensation effect is shown to be due to the existence of 
the chiral glass phase.
We demonstrate, in agreement with experiments, that
this effect may be
present in the ceramic superconductors which show the paramagnetic 
Meissner effect.
{}~\\
{}~
\end{minipage}
}}

\maketitle



One of the most fascinating discoveries in condensed matter physics
is the paramagnetic Meissner effect (PME) in certain ceramic superconductors
\cite{Sved,Braunish}. The nature of the unusual paramagnetic behaviour may be
related to the appearance of spontaneous suppercurrents (or of orbital moments)
\cite{Kusmartsev}. The latter appear due to the existence
of $\pi$-junctions characterized by the negative Josephson couplings
\cite{Kusmartsev,Bulaevskii}. 
Furthermore, Sigrist and Rice argued that the PME in the high-$T_c$ 
superconductors is consistent with the $d$-wave superconductivity\cite{Sirgist}.
This effect is succesfully reproduced in a single
loop model\cite{Sirgist} as well as in a model of interacting junction-loops
\cite{Dominguez,KawLi}.

The mechanism of the PME based on the $d$-wave symmetry of the order parameter
remains ambiguous because it is not clear why this effect could not be
observed in many ceramic materials. More importantly, the paramagnetic response
has been seen even in the conventional Nb \cite{Thompson,Kostic,Pust} 
and Al \cite{Geim} superconductors.
In order to explain the PME in terms of conventional superconductivity
one can employ the idea of the flux compression inside of a sample. Such
phenomenon becomes possible in the presence of the inhomogeneities\cite{Larkin}
or of the sample boundary\cite{Moshchalkov}.
Thus the intrinsic mechanism leading to the PME is still under 
debate\cite{Geim,Sigrist1}.

Recently Heinzel {\em et al.}\cite{Heinzel} have shown that the PME may 
be analyzed by the compensation technique based on the measurement of the
second harmonics of the magnetic ac susceptibility.
Their key observation is that the so called compensation effect (CE)
appears only in the 
samples which show the PME but not in those which do not.
Overall, this effect  may be detected in the following way.
The sample is cooled in
the external dc field down to a low temperature and then the field is
switched off. At the fixed low $T$ the second harmonics are monitored 
by applying the dc and ac
fields to the sample.
Due to the presence of non-zero spontaneous orbital
moments the remanent magnetization or, equivalently, the internal field appears
in the cooling process.
If the direction of the external dc field is 
identical to that during the field cooled (FC) procedure, the induced  
shielding
currents will reduce the remanence. Consequently, the absolute value
of the second harmonics $|\chi_2|$ decreases until 
the signal of the second harmonics is minimized
at a field $H_{dc}=H_{com}$.
Thus the CE is a phenomenon in which the external and internal fields are
compensated and the second harmonics become zero.

The goal of this paper is to explain the CE theoretically by 
Monte Carlo simulations.
Our starting point is based on the possible existence of the chiral glass
phase\cite{Kawamura} in which the remanence necessary for observing the CE 
should occur in the cooling procedure.
Such remanence phenomenon is similar to what happens in spin glass.
Furthermore, the PME related to the CE can also be observed in the chiral glass
phase\cite{KawLi,Kawamura,KawLi1}. There are several experimental 
results\cite{Matsuura} which appear to corroborate the existence of such a 
novel glassy phase in ceramic high-$T_c$ superconductors.  

In the chiral glass phase
the frustration due to existence of 0- and $\pi$-junctions
(0-junctions correspond to positive Josephson
contact energies) leads to non-zero supercurrents\cite{Kawamura}. 
The internal field (or the remanent 
magnetization) induced by the supercurrents
in the cooling process from high temperatures to
the chiral glass phase may compensate the external dc field.

We model  ceramic superconductors by the three-dimensional
XY model of the Josephson network 
with  finite self-inductance.
We show that in the FC regime the CE appears 
in the samples 
which show the PME but not 
in  those containing only $0$-junctions.
In the zero field cooled (ZFC) regime decreasing the external dc field
also gives rise to the CE in the frustrated 
ceramics. 
Both of these findings
agree with the experimental data of Heinzel {\em et al}\cite{Heinzel}.

We neglect the charging effects of the grain and
consider the following Hamiltonian\cite{Dominguez,KawLi}
\begin{eqnarray}
{\cal H} = - \sum _{<ij>} J_{ij}\cos (\theta _i-\theta _j-A_{ij})+ \nonumber\\
\frac {1}{2{\cal L}} \sum _p (\Phi_p - \Phi_p^{ext})^2, \nonumber\\
\Phi_p \; \; = \; \; \frac{\phi_0}{2\pi} \sum_{<ij>}^{p} A_{ij} \; , \;
A_{ij} \; = \; \frac{2\pi}{\phi_0} \int_{i}^{j} \, \vec{A}(\vec{r}) 
d\vec{r} \; \; ,
\end{eqnarray}
where $\theta _i$ is the phase of the condensate of the grain
at the $i$-th site of a simple cubic lattice,
$\vec A$ is the fluctuating gauge potential at each link
of the lattice,
$\phi _0$ denotes the flux quantum, 
$J_{ij}$ denotes the Josephson coupling
between the $i$-th and $j$-th grains, 
${\cal L}$ is the self-inductance of a loop (an elementary plaquette),
while the mutual inductance between different loops
is neglected.
The first sum is taken over all nearest-neighbor pairs and the
second sum is taken over all elementary plaquettes on the lattice.
Fluctuating  variables to be summed over are the phase variables,
$\theta _i$, at each site and the gauge variables, $A_{ij}$, at each
link. $\Phi_p$ is the total magnetic flux threading through the 
$p$-th plaquette, whereas $\Phi_p^{ext}$ is the flux due to an 
external magnetic field applied along the $z$-direction,
\begin{equation}
\Phi_p^{ext} = \left\{ \begin{array}{ll}
                   HS \; \;  & \mbox{if $p$ is on the $<xy>$ plane}\\
                   0  & \mbox{otherwise} \; \; ,
                        \end{array}
                  \right. 
\end{equation}
where $S$ denotes the area of an elementary plaquette.
The external field $H$ includes the dc and ac parts
and it is given by
\begin{equation}
H \; \; = \; \; H_{dc} + H_{ac} \cos(\omega t) \; \; .
\end{equation}
It should be noted that the dc field is necessary to generate even harmonics.


In the present paper, we consider two models with two types of 
bond distributions.
Model I: the sign of the Josephson couplings
could be either positive
(0-junction) or negative ($\pi$-junction) and
the spin glass type bimodal ($\pm J$) distribution
of $J_{ij}$ is taken.
The coexistence of 0- and $\pi$-junctions gives rise to frustration
even in zero external field and
the chiral glass phase may occur at low
temperatures\cite{Kawamura}. Model II: 
the interactions $J_{ij}$ are assumed to be 'ferromagnetic' and
distributed uniformly between 0 and 2$J$. Obviously,
there is no
frustration in zero external field in this model.
It has been also demonstrated that the PME is present in model I but not in
model II\cite{KawLi}.

The ac linear susceptibilty of models I and II
has been studied\cite{KawLi} by Monte Carlo simulations. 
It was found that, due to the frustration, model I exhibits 
much stronger dissipation than model II in the low frequency regime.
Here we go beyond our previous calculations of the linear ac susceptibility
\cite{KawLi}. We study the dependence of the second harmonics
as a function of the dc field. In this way, we can make a direct comparison
with the CE observed in the experiments \cite{Heinzel}.
The second harmonics of a similar Josephson network model with a finite 
self-inductance were considered by Wolf and Majhofer\cite{Majhofer}. However,
these authors dealt with the two-dimensional version of model II
and the CE has not been studied. In this paper we are mainly interested in
the CE in the frustrated three-dimensional system described by model I. 

\begin{figure}
\epsfxsize=3.2in
\centerline{\epsffile{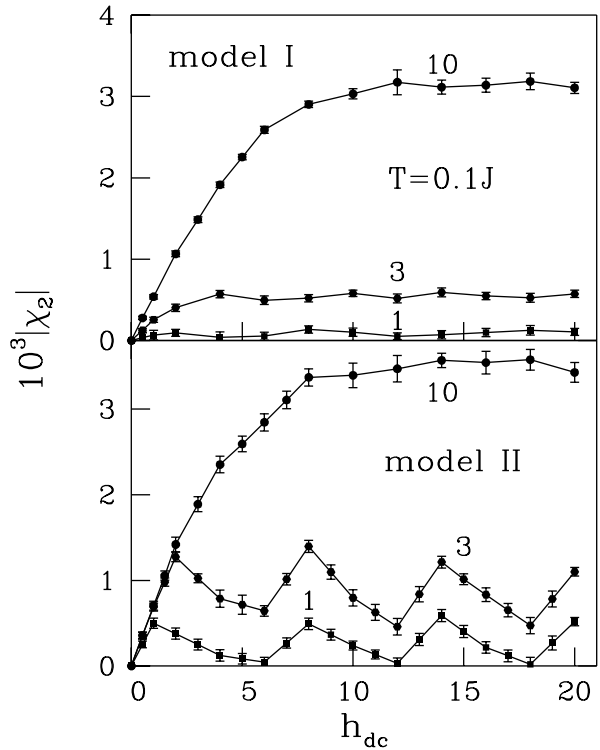}}
\caption{
The dependence of $|\chi_2|$ on $h_{dc}$ for model I and
model II at $T=0.1$. The values of $\tilde{{\cal L}}$ are
chosen to be equal to 1, 3 and 10 as shown next to the curves.
The results are averaged over 20 samples.}
\end{figure}

The dimensionless magnetization along the $z$-axis mormalized per plaquette,
$\tilde{m}$, is given by
\begin{equation}
\tilde{m} \; \; = \; \; \frac{1}{N_p\phi_0} \; \sum_{p<xy>} 
(\Phi_p - \Phi_p^{ext}) \; \; ,
\end{equation}
where the sum is taken over all $N_p$ plaquettes on the $<xy>$ plane of the
lattice. The real and imaginary parts of the ac second order susceptibility
$\chi'_2(\omega)$ and $\chi''_2(\omega)$ are calculated as
\begin{eqnarray}
\chi'_2(\omega) \; \; &=& \; \; \frac{1}{\pi h_{ac}}
\int_{-\pi}^{\pi} \; \tilde{m}(t) \cos(2\omega t)d(\omega t) \; \; , \nonumber\\
\chi''_2(\omega) \; \; &=& \; \; \frac{1}{\pi h_{ac}}
\int_{-\pi}^{\pi} \; \tilde{m}(t) \sin(2\omega t)d(\omega t) \; \; ,
\end{eqnarray}
where $t$ denotes the Monte Carlo time. The dimensionless ac field 
$h_{ac}$, dc field $h_{dc}$ and inductance 
$\tilde{\cal L}$ are defined 
as follows 
\begin{eqnarray}
h_{ac} \; \; = \; \; \frac{2\pi H_{ac}S}{\phi_0} \; \; , \; \;
h_{dc} \; \; = \; \; \frac{2\pi H_{dc}S}{\phi_0} \; \; , \; \;\nonumber\\
\tilde {\cal L} \; \; = \; \; (2\pi /\phi_0)^2 J{\cal L}.
\end{eqnarray}
The dependence of $\tilde{\cal L}$ on the parameters of the system
such as the critical current and the typical size of the grains is
discussed in \cite{KawLi,KawLi1}.

Our results have been obtained by employing 
Monte Carlo simulations based on the standard Metropolis updating 
technique.
While Monte Carlo simulations involve
no real dynamics, one can still expect that they give useful information on
the long-time behavior of the system. In fact, the amplitude of the ac field
we use is much smaller than the typical energy of the dc part.
On the other hand, the characteristic time for the sintered samples, 
which are believed to be captured
by our model, is of order $10^{-12} s$\cite{Majhofer}. This time has the same 
order of magnitude as a single Monte Carlo step. So 
the period of oscillations chosen in the present work is much longer than
the characteristic time (see below).
For such a weak and slowly changing ac field the system can be regarded 
as being in
quasi-equilibrium and the Monte Carlo updating may be applied.
A priori, the validity of this approximation is
not clear but it may be justified by comparing
our results with those obtained by other approaches to the 
dynamics such as considered in ref. \cite{Majhofer}.
For the first harmonics, our method and the method of ref. \cite{Majhofer}
yield results that agree qualitatively.
Furthermore, our results presented in Fig.1 for
the second harmonics are also in a qualitative
agreement with the corresponding results obtained by 
solving the equations of motion \cite{Majhofer}.
So one can expect that the standard Monte Carlo may actually give 
reasonable results for the CE.

\begin{figure}
\epsfxsize=3.2in
\centerline{\epsffile{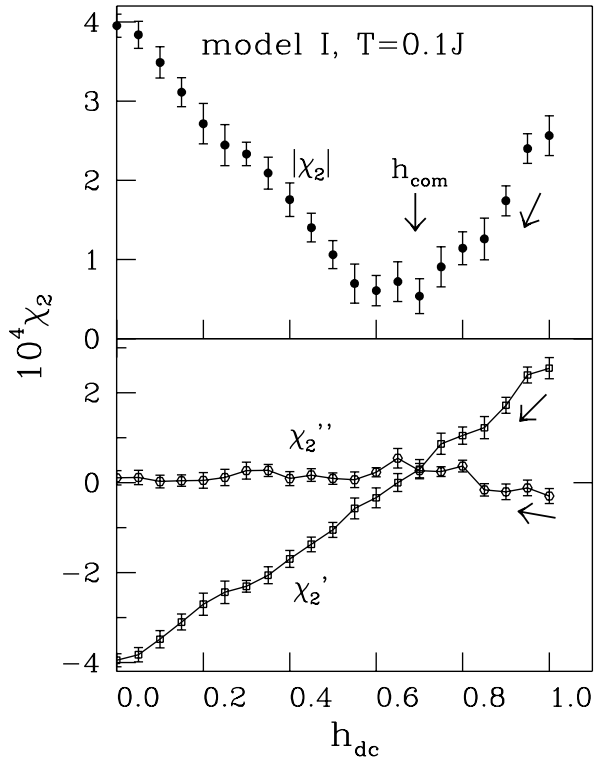}}
\caption{
The second harmonics of model I obtained
after field cooling in a dc field $h_{dc}=1$ from $T=0.7$ to $T=0.1$.
The temperature is reduced in steps of 0.05.
At the lowest $T=0.1$ the dc field used in cooling is switched off
and the second harmonics are generated by applying the combined
field (3). The $dc$ field is stepwise reduced from $h_dc=1$ to $h_dc=0$.
The inductance is chosen to be equal to $\tilde{{\cal L}}=4$.
The arrows indicate the sense of the changes in the
dc field. The results are averaged over 40 samples and are
qualitatively the same as those presented in Fig. 1 of Ref. [15].}
\end{figure}

We choose the gauge where the bond variables $A_{ij}$ 
along the $z$-direction are fixed to be zero. The lattice studied are simple
cubic with $L\times L\times L$ sites and free boundary conditions  are
adopted. In all calculations presented below, we take $L=8$ and 
$\omega=0.001$. The sample average is taken over 20-40 independent bond
realizations. $\chi_2(\omega)$ has been estimated following the procedure
in\cite{KawLi,Hung}.
Namely, at the beginnig of a given Monte Carlo run, we first 
switch on the field (3). Then, after waiting for initial $t_0$ Monte Carlo
steps per spin (MCS), we start to monitor the time variation of 
the magnetization, $t_0$ is being chosen so that all transient phenomena 
can be considered extinct. We set $t_0$ to be $2\times 10^4$ MCS. After 
passing the point $t=t_0$, $\tilde{m}(t)$ is averaged over typically 200
periods, each period contains $t_T$ MCS ($t_T=2\pi/\omega$). The
real and imaginary parts of the second order ac susceptibility are then
extracted via Eq. (5).
We set $h_{ac}=0.1$, corresponding to $\approx 0.016$ flux quantum per
plaquette. Smaller value of $h_{ac}$ turned out to leave the results 
almost unchanged.

\begin{figure}
\epsfxsize=3.2in
\centerline{\epsffile{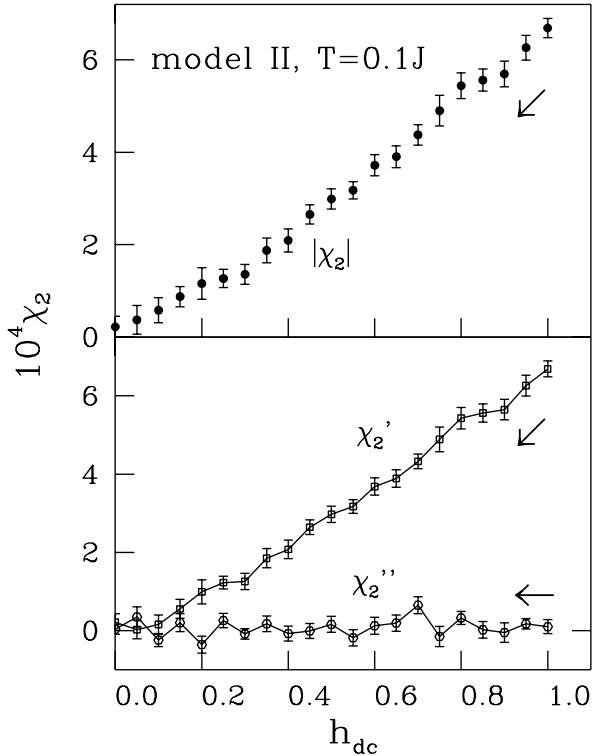}}
\caption{
The procedure to generate the second harmonics is the same
as in Fig. 2 but for model II. The system is cooled
in a dc field $h_{dc}=1$ from $T=1.4$ to $T=0.1$.
The results are averaged over 20 samples.}
\end{figure}

The dependence of $|\chi_2|$, $|\chi_2|=\sqrt{(\chi'_2)^2 + (\chi''_2)^2}$,
on $h_{dc}$ at $T=0.1J$ is presented in Fig.1. For small values of 
$\tilde{{\cal L}}$, the oscillation of $|\chi_2|$  shows up. 
Such oscillation has 
been found for the two-dimensional superconductors in 
Ref. \cite{Majhofer} and its nature is related to the lattice periodicity. 
Our new observation is that the oscillatory behavior is 
still present in the superconductors with $0$- and $\pi$-junctions
(model I) but to less extent compared to
model II. It is clear from Fig. 1 that 
$|\chi_2|$  does not decrease at large $h_{dc}$ but gets saturated. This is
an artifact of the assumption that the Josephson contact energies
$J_{ij}$ are field-independent. The field dependence
of $J_{ij}$ should remove the saturation of $|\chi_2|$ at strong dc 
fields\cite{Majhofer,Ji}.

In order to study the difference between model I and model II
through the CE we have to consider the weak field region where the PME 
may be observed. For model I the PME appears clearly 
for $h_{dc} \leq 1$ \cite{KawLi}. So the largest $h_{dc}$
we take is 1. In this weak field regime there is no
periodicity of $|\chi_2|$ versus $h_{dc}$ which may complicate the study of the
CE. The chiral glass phase 
is found to exist below a critical value of the inductance 
$\tilde{{\cal L}}_c$ where $5 \leq \tilde{\cal{L}}_c \leq 7$ \cite{Kawamura}.
One has to choose, therefore, an $\tilde{{\cal L}}$ which is smaller than 
its critical value and 
in what follows we take $\tilde{{\cal L}}=4$. 

In this paper we focus on the system size $L=8$, $\tilde{{\cal L}}=4$,
$\omega=0.001$, and $T=0.1$. Our preliminary studies show that the
qualitative results do not depend on the choise of the parameters of
the system.

Fig. 2 shows the dependence of second harmonics on $h_{dc}$ in the
FC regime for the superconductors described by model I. 
Our calculations follow exactly
the experimental procedure of Heinzel {\em et al}\cite{Heinzel}.
First the system is cooled in the dc field $h_{dc}=1$
from $T=0.7$ down to $T=0.1$ which is below
the paramagnet-chiral glass transition temperature $T_c\approx 0.17$
\cite{Kawamura}. The temperature step is chosen to be equal to 0.05. At each 
temperature, the system is evolved through
 2$\times 10^4$ Monte Carlo steps. When the lowest 
temperature is reached the dc field used in cooling is switched off and 
we apply the combined field given by Eq. (3). We monitor the second harmonics
reducing the dc field from $h_{dc}=1$ to zero stepwise by
an amount of $\Delta h_{dc}=0.05$. $|\chi_2|$ reaches minimum at
the compensation field $h_{com}=0.7\pm 0.05$. At this point,
similar to the experimental findings\cite{Heinzel},
the intersection of $\chi'_2$ and $\chi''_2$ is observed. This fact 
indicates that at $H_{com}$ the system is really in the compensated state.
Furthermore, in accord with the experiments, at the compensation point the 
real and imaginary parts should change their sign\cite{Heinzel}.
Our results show that $\chi'_2$ changes its sign roughly at $h_{dc}=h_{com}$.
A similar behavior is also displayed by $\chi''_2$
but it is harder to observe due to a smaller amplitude of $\chi''_2$.

Fig. 3 shows the dependence of the second harmonics on $h_{dc}$ in FC
regime for model II. The calculations are carried out in 
the same way as for model I. A difference is that we start to
cool the system from $T=1.4$ which is above the superconducting transition
point $T_s\approx 0.9$ ($T_s$ is estimated from the maximum of the specific
heat\cite{KawLi} for $\tilde{{\cal L}}=4$ and the results are 
not shown here). The temperature step
is set equal to 0.1. Obviously, $|\chi_2|$ decreases with
decreasing $h_{dc}$ monotonically. Thus, there is no CE because
the remanent magnetization does not appear in the cooling process.
This result is again in accord with the experimental data\cite{Heinzel}.

\begin{figure}
\epsfxsize=3.2in
\centerline{\epsffile{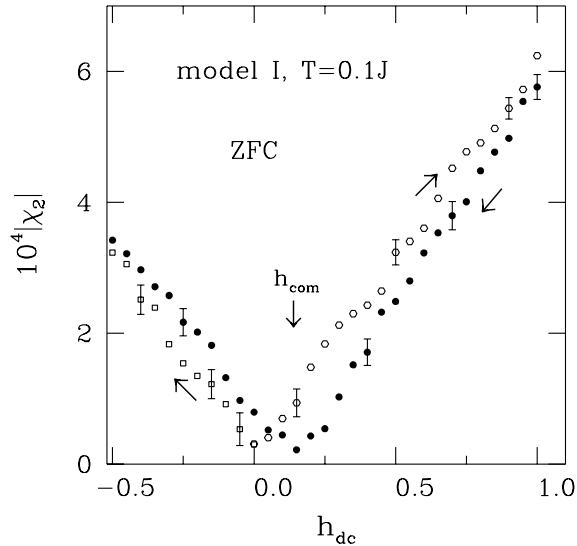}}
\caption{
The dependence of $|\chi_2|$ on $h_{dc}$ obtained in the ZFC regime for
model I.
The solid circles correspond to the case
when the dc field is decreased  from $h_{dc}=1$
to -0.5.  The open hexagons and squares correspond
to the increase of $h_{dc}$ from
zero to 1 and to its decrease from zero to -0.5, respectively.
The inductance is chosen to be equal to $\tilde{{\cal L}}=4$.
The sense of changes of the dc field is marked by the arrows.
The results are averaged over 25 samples. 
The compensation field in the case when the field is decreased
is $h_{com}=0.15\pm 0.05$.}
\end{figure}

We now turn to the ZFC regime. The experiments\cite{Heinzel}
show that no CE can be expected if after the ZFC procedure one increases the
dc field. However, if the field is decreased  a remanent magnetization is 
developed and the CE appears\cite{Heinzel}. 
The results of our simulations for the
ceramic superconductors described by model I are shown in Fig. 4. 
As in the FC regime the system
is cooled from $T=0.7$ to $T=0.1$ but without the external field. Then
at $T=0.1$ we apply
the field given by Eq. (3) and study three cases. In one of them
$h_{dc}$ is decreased from $h_{dc}=1$ to -0.5. The values of 
$|\chi_2|$ are represented by solid circles in Fig. 4.
The CE is clearly seen at $h_{com}=0.15\pm 0.05$. At this point the real
and imaginary parts of the second harmonics also intersect (the results are not
shown). It is not surprising that the $h_{com}$ in 
the ZFC regime appears to be smaller 
than in the FC regime. Fig. 4 shows also the dependence of $|\chi_2|$ on
the dc field  when it changes
from $h_{dc}=0$ to 1 (open hexagons) and from $h_{dc}=0$ to -0.5 (open squares).   
Obviously, no CE is observed in this case. The results presented in 
Fig. 4 qualitatively
agree with those shown in Fig.2 of Ref.\cite{Heinzel}. 

In conclusion we have shown that the CE may be explained, at least 
qualitatively,
by using the chiral glass picture of the ceramic superconductors.
The CE is shown to appear in the chiral glass phase in which the PME is
present but not in the samples without the PME.


\noindent

We thank M. Cieplak for a critical reading of manuscript and H. Kawamura,
D. Dominguez, A. Majhofer and S. Shenoy for discussions. Financial support from the Polish agency KBN
(Grant number 2P03B-025-13) is acknowledged. 

\par


\noindent

\begin{references}
%
\bibitem{Sved} P. Svedlindh, K. Niskanen, P. Nordblad, L. Lundgren,
B. L\"onnberg and T. Lundstr\"om, 
Physica C {\bf 162-164}  (1989) 1365.
\bibitem{Braunish}  W. Braunish, N. Knauf, V. Kataev, S. Neuhausen, A. Grutz, A. Kock,
B. Roden, D. Khomskii, and D. Wollleben, Phys. Rev. Lett. {\bf 68},
1908 (1992);
W. Braunisch, N. Knauf, G. Bauer, A. Kock, A. Becker, B. Freitag, A. Gr\"utz,
V. Kataev, S. Neuhausen, B. Roden, D. Khomskii, D. Wohlleben, J. Bock and 
E. Preisler, Phys. Rev. B {\bf 48}  (1993) 4030.
\bibitem{Kusmartsev} F.V. Kusmartsev, Phys. Rev. Lett. {\bf 69}  (1992) 2268;
J. of Superconductivity {\bf 5}  (1992) 463.
\bibitem{Bulaevskii} L. N. Bulaevskii, V. V. Kuzii, and A. A. Sobyanin,
JETP Lett. {\bf 25}, 290 (1977); V. B. Geshkenbein, A. I. Larkin, 
and A. Barone, Phys. Rev. B {\bf 36}, 235 (1987);
B. I. Spivak and S. A. Kivelson, Phys. Rev. B {\bf 43}, 3740 (1991).
\bibitem{Sirgist} M. Sigrist and T.M. Rice,
J. Phys. Soc. Jpn. {\bf 61}  (1992) 4283; Rev. Mod. Phys.
{\bf 67} (1995) 503
\bibitem{Dominguez} D. Dom\'inguez, E.A. Jagla and C.A. Balseiro,
Phys. Rev. Lett. {\bf 72}  (1994) 2773.
\bibitem{KawLi} H. Kawamura and M.S. Li, Phys. Rev. B
{\bf 54} (1996) 619.
\bibitem{Thompson}
D. J. Thompson, M. S. M. Minhaj, L. E. Wegner, and J. T. Chen,
Phys. Rev. Lett. {\bf 75}, 529 (1995).
\bibitem{Kostic} P. Kostic, B. Veal, A. P. Paulikas, U. Welp,
V. R. Todt, C. Gu, U. Geiser, J. M. Williams, K. D. Carlson, and
R. A. Klemm, Phys. Rev. B {\bf 53}, 791 (1996); {\bf 55}, 14649 (1997).
\bibitem{Pust}L. Pust, L. E. Wegner, and M. R. Koblischka,
Phys. Rev. B {\bf 58}, 14191 (1998).
\bibitem{Geim}A. K. Geim, S. V. Dubonos, J. G. S. Lok, M. Henini,
and J. C. Maan, Nature {\bf 396}, 144 (1998).
\bibitem{Larkin} A. E. Koshelev and A. I. Larkin, Phys. Rev. B {\bf 52},
13559 (1995); A. E. Khalil, Phys. Rev. B {\bf 55}, 6625 (1997).
\bibitem{Moshchalkov} V. V. Moshchalkov, X. G. Qui, and V. Bruyndoncz,
Phys. Rev. B {\bf 55}, 11793 (1997).
\bibitem{Sigrist1} M. Sigrist, Nature {\bf 396}, 110 (1998).
\bibitem{Heinzel} Ch. Heinzel, Th. Theilig, and P. Ziemann,
Phys. Rev. B {\bf 48}, 3445 (1993).
\bibitem{Kawamura} H. Kawamura and M. S. Li, Phys. Rev. Lett.
{\bf 78}, 1556 (1997).
\bibitem{KawLi1}  H. Kawamura and M. S. Li,
J. Phys. Soc. Jpn {\bf 66}, 2110 (1997).
\bibitem{Matsuura} L. Leylekian, M. Ocio and J. Hammann, Physica C 
{\bf 185-189}, 2243 (1991); Physica B {\bf 194-196}, 1865 (1994);
M. Matsuura, M. Kawachi, K. Miyoshi, M. Hagiwara, and  K. Koyama,
J. Phys.Soc. Jpn {\bf 64},, 4540 (1995).
\bibitem{Majhofer} T. Wolf and A. Majhofer, Phys. Rev. B {\bf 47}, 5383 
(1993).
\bibitem{Hung} M. Z. Cieplak, T. R. Gawron, and M. Cieplak, Phys. Rev. B
{\bf 39}, 6757 (1989); M. S. Li, T. Q. Hung, and M. Cieplak,
J. Phys. (France) I {\bf 6}, 1 (1996)
\bibitem{Ji} L. Ji, R. H. Sohn, G. C. Spalding, C. J. Lobb, and M. Tinkham,
Phys. Rev. B {\bf 40}, 10936 (1989)






%
\end{references}
\end{document}